\documentclass[preprint2]{aastex63}

\shorttitle{Formation of the CCKB}
\shortauthors{R. Gomes}

\begin{document}


\title{The Formation of the Cold Classical Kuiper Belt by a Short Range Transport Mechanism}

\author{Rodney Gomes}
\affil{Observat\'orio Nacional \\
Rua General Jos\'e Cristino 77, CEP 20921-400, Rio de Janeiro, RJ,  Brazil}
\email{rodney@on.br}

\begin{abstract} 
The Classical Kuiper Belt is populated by a group of objects with low inclination orbits, reddish colors
and usually belonging to a binary system. This so called Cold Classical Kuiper Belt is considered
to have been formed in situ from primordial ice pebbles that coagulated into planetesimals hundreds of kilometers
in diameter. According to this scenario, the accretion of pebbles into large planetesimals
would have occurred through the streaming instability mechanism
that would be effective in the primordial Solar System disk of gas
and solids. Nevertheless other objects with the same color characteristics as those found in the Cold Classical
Kuiper Belt can be encountered also past the 2:1 mean motion resonance with Neptune as scattered or detached objects.
Here I propose a mechanism that can account for both the cold Classical Kuiper Belt objects and other reddish
objects outside the Classical Kuiper Belt. According to the proposed
scenario, reddish objects were primordially in the outer portion of the planetesimal disk which
was however truncated somewhere below $\sim 42$ au. In this manner the cold Classical Kuiper Belt and 
its scattered / detached counterpart were respectively transported outwards by a short range or slightly scattered
to their present locations. Resonant objects were also formed by the same process. 
This mechanism is aimed at explaining the distribution of all objects that share
the same color characteristics as coming from a common origin in the outer borders of the 
primordial planetesimal disk. According to the scenario here proposed the Cold Classical Kuiper Belt would have been
formed $\sim 4$ au inside its present location with a total mass $20-100$ times as large as its present value.

\end{abstract}

%\begin{graphicalabstract} \includegraphics{figs/grabs.pdf}

%\begin{keywords}
%Kuiper Belt  Planetary Migration
%\end{keywords}

%\maketitle

\section{Introduction}

Trans-Neptunian objects (TNOs) show a conspicuous orbital architecture that is
best characterized through the classification of the total population into
specific subpopulations known as: the classical Kuiper belt (CKB), cold (CCKB)
and hot, the resonant population, the scattered population and the detached
population. This remarkable orbital distribution into several subpopulations is
a consequence of the gravitational interaction of the giant planets and a disk of
planetesimals in the early Solar System that accounted for the implantation of the planets on
their present orbits and the planetesimals that induced the planets
migration onto their several different populations. That process
included resonance trapping and evolution in resonance of the planetesimals
with the major planets (mainly Neptune) \citep{malhotra1995, gomes2000}, scattering by close encounters with
the migrating planets, resonance sticking of scattering planetesimals into mean motion 
resonances (MMR) 
and Kozai resonances with Neptune and possible escape from these resonances \citep{gomes2003, gomes2011, lykawka-mukai2007}.
Among these TNOs subpopulations, the CCKB is presently regarded as the set of
objects that experienced the least interactions with the migrating planets.

The CCKB is possibly the part of the Kuiper belt that best represents the TNOs as first conceived by its
first proposers \citep{Kuiper1951, Edgeworth1949}. It is roughly a set of objects with relatively low
eccentricity and low inclination orbits and semimajor axes roughly between 42 au
and 47 au. Nevertheless, we cannot claim that the eccentricities of the CCKB is
as low as expected if these objects were perturbed during the age of the Solar
System solely by the planets at their present orbital configurations. It is thus
understood that the migrating planets had some influence albeit small on the present CCKB orbits
 \citep{nesvorny2015, dmc2012, gomesetal2018, ribeiroetal2019}. This
perturbation on the CCKB's  objects must have been weak enough in order to avoid not only an excessive
excitation of the CCKB orbits but also to preserve their binary feature. Close
approach perturbations with the migrating planets would destroy most binary
systems which is contrary to observations \citep{parker-kavelaars2010}.

In a not distant past, there have been good reasons to suppose that the CCKB was
transported from inner regions of the primordial planetesimal disk by some
mechanism associated with planetary migration \citep{levmorb2003, levisonetal2008}.
The reason for this was basically due to the fact that the CCKB has
presently a very low mass  estimated from $3 \times 10^{-4} M_{\oplus}$ \citep{fraser-a-2014} to
$10$ times as that number \citep{nesvornyetal2020}. Classical
planetesimal accretion theories would demand a fairly high initial mass in the
CKB region in order to create objects as large as the real ones.
Yet collisions among these objects are not efficient enough
to grind them to the present CCKB mass during the
Solar System age \citep{kenyonluu1999}. Moreover, a large mass 
in the primordial planetesimal disk would push Neptune to the
outer border of the disk where the present KB is today \citep{gomesetal2004}. More recently, however, new
theories on planetesimal accretion, induced by streaming instability \citep{Drazkowska-Dullemond2014, johansenetal2007,youdin-goodman2005}
(see also \cite{morb-nesv2020} for a review concerning the formation of the Kuiper Belt)
can explain the formation of large planetesimals ($100$ km diameter) directly
from cm-sized ice pebbles. The main idea behind these mechanisms is based 
on the radial drift of
these pebbles rotating in a sub-keplerian regime due to drag induced by the disk gas. As the pebbles
drift inwards, they accumulate into inner regions of the disk where the
drift decreases or stalls due to too much mass in the pebbles that force the gas to
rotate with the pebbles frequency. This on its turn provokes the
accumulation of pebbles into certain regions of the disk which on its turn induces a
gravitational instability that makes the pebbles to accumulate into a single large
object. Streaming stability has been shown also to yield planetesimals in orbits that are compatible with
the spatial orientation of Kuiper Belt binary orbits \citep{nesvornyetal2019}. 

A classical definition of the CCKB comes from its dynamical properties. 
According to that definition the CCKB is formed by objects in a certain
semimajor axis range ($\sim 42-47$ au) and inclinations below a certain value
(usually $4^{\circ}$ or $5^{\circ}$). But CCKB objects also differ from the rest of the
TNOs by their physical properties of which their colors are probably the main
characteristic. In fact, CCKB objects are mostly reddish, opposed to other TNOs whose color range
from red to neutral colors. Another important particular characteristic of CCKB objects is
that they  usually appear in pairs, since most of them belong to a binary system. 
Nevertheless the binary TNOs inventory includes mostly cubewanos with very few 
scattering or detached objects, still fewer when we consider only similar size members binaries 
which are mostly associated to the CCKB objects. 

With this
in mind, I propose to define the Extended Cold Population (ECP) \footnote{I keep the 'cold' adjective
for this population's nomenclature to remind that all these objects share a common origin as that of the CCKB, 
acknowledging however that this extended population will include orbits with
much higher orbital eccentricities and inclinations than those in the CCKB} as formed by objects in a certain range of
color indices, like B-R.
With this definition the orbital distribution of the ECP objects
invades a larger region than that defined above based only in orbital
characteristics. 
The idea is
that the ECP is distinguished from the rest of the TNOs by the
primordial region where it was formed and some primordial mechanism placed them
not only on its classical KB region but some of them were transported to regions
outside the CCKB. The main proposal in this paper
is that the CCKB was not really formed in situ but
experienced a short transportation process from the outermost part of the original planetesimal disk
which would have an outer border somewhere around $42$ au.
My second proposal is that this assumption allows for the explanation of reddish objects outside the CCKB as well as 
the formation of the CCKB objects in a more massive environment than the present CCKB since
although streaming instability theories may allow for the construction of large objects like those
of the CCKB, it is however questionable whether such a low mass as $3 \times 10^{-4}$ to $3 \times 10^{-3} m_{\oplus}$
can be formed within that theory.

This paper is divided as follows. In Section 2, I define the ECP that will
serve as a reference to search for the best model to produce it. Section 3 describes the
model used in this work. In Section 4, the main results are presented. Section 5 explains the mechanisms that were effective in generating
the orbits associated to the ECP. In Section 6, I present some examples of numerical simulations
including four or five planets and a disk of perturbing planetesimals in order to check for
the viability of the proposed model. In Section 7, I discuss the results and present conclusions.

\section{The Extended Cold Population}

As advanced in the Introduction, defining
the cold population just by its orbit characteristics may not be the best
approach. We might instead define the cold population by their colors or their binary feature. Of
course, none of these possible definitions completely characterizes the cold population. But
defining the cold population solely by its orbital characteristic can hide
important information as to its origin. The inventory of discovered binary systems
among the TNOs is not so large (in particular with respect to distant TNOs) 
 as that of objects with defined color characteristics, thus I choose TNOs colors as
a constraint to define the ECP.

\begin{figure} 
\centering 
\includegraphics[scale=.57]{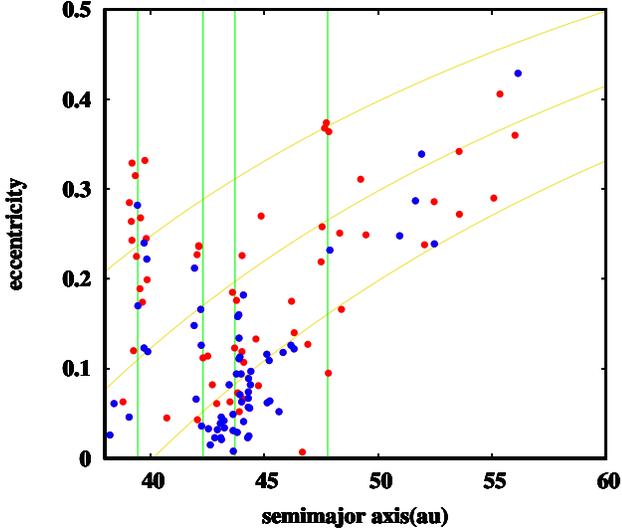}
\caption{Distribution of semimajor axes and eccentricities of observed objects
belonging to the ECP. Blue circles stand for objects with the orbital inclination
$I<4^{\circ}$. 
Otherwise they are represented by red circles}
\label{fig-br} 
\end{figure}

In Johnston's archive website \footnote{http://www.johnstonsarchive.net/astro/tnoslist.html} one finds orbital elements of
TNOs and their color characteristics, such as their B-R magnitude and their
taxonomical color type (RR,IR,BR, etc.). I could define the ECP either by
including in it all the objects belonging to the most reddish taxonomical types
(RR or RR+IR) or by fixing a B-R magnitude above which all objects would
belong to the ECP. To decide on the best choice, I separated a constrained part
of the TNOs from the above table, defined by ranges in $a$ (semimajor axes) and $q$ (perihelion distances) given by
$42.5$ au $< a < 46.5$ au and $q>38$ au, which will stand for the CCKB defined in this work.
Then I computed the ratio of the number of objects that have their orbital inclination $I<4^{\circ}$ to the total
number of objects in that subpopulation. This was done for objects associated
to the RR type, the RR+IR type and those that have B-R magnitude greater than
$1.58$ \footnote{This magnitude was chosen based on a balance between yielding a high
ratio of low eccentricity members to all others and keeping a reasonable number
of objects in the ECP}. The largest ratio was obtained for the last
subpopulation, the one defined by magnitude B-R $> 1.58$. This is a way of defining
the ECP primarily by its color characteristics but somehow constrained by the
fact that the CCKB is an important part of the total ECP and that it is composed mostly of low 
inclination objects. 

Figure \ref{fig-br} shows the distribution of semimajor axes and eccentricities for
objects that have their B-R magnitude larger than $1.58$, taken from Johnston's archive website.
This will be defined as
the observed ECP. To best compare with the results shown in following sections I farther constrain
the ECP just for objects with $37$ au $< a < 60 $ au. 
I define three different subpopulations in the ECP: the CCKB, defined as
objects with $42.5 $ au $ < a < 46.5$ au and $q > 38$ au; the distant cold population (DCP), defined by
semimajor axes above the 2:1 MMR with Neptune ($a > 49 $ au) and below $60 $au, this upper limit
chosen since there are few objects with B-R magnitude greater than $1.58$ and $a > 60$ au; the resonant population, defined for semi
major axes around the nominal semi
major axis for a specific MMR with Neptune. For the case of the resonances 5:3 and 7:4 objects with $q>38$ are considered belonging to the CCKB as stated above. For the 2:1 resonance I consider objects with their semimajor axes $1$ au above or below the nominal resonant semimajor axis as belonging to the 2:1 resonance. Motivated by Fig. \ref{fig-br}, I consider only
the resonances 3:2, 5:3, 7:4 and 2:1. The DCP is specially revealing since it shows scattered / detached orbits
with fairly low inclinations. Three of them, 2001 UR163, 1995 TL8 and  2002 GZ31, are still more conspicuous.
They have semimajor axes in the range $50.9 $ au $ < a < 52.5$ au, perihelion distances in the range $36.8$ au $ < q < 39.9$ au
and inclinations below $1.1^{\circ}$. Far from important MMR with Neptune  and with relatively low perihelia and low
inclinations, their orbits are hardly explainable by classical mechanisms including resonance sticking followed by
the coupling of MMR with Neptune with
Kozai resonance \citep{Brasiletal2014, Brasiletal2014b, gomes2011}.

\section{The Model}

The simulations undertaken for this work were initially motivated aiming at
creating a transportation mechanism that would preserve the planetesimals'
eccentricity. Although a local origin for the CCKB is widely accepted as the
best explanation for the present orbital distribution of the CCKB objects, it
may however be difficult to accommodate a local formation scenario that can
account for both the CCKB and the rest of the ECP, in particular the DCP. 
Transporting mechanisms for the CCKB have failed mostly because they
yielded too excited orbits for the CCKB objects \citep{levisonetal2008, levmorb2003}.
This is basically due to the eccentricity excitation as an
effect of the resonance sweeping mechanism to transport planetesimals from
inner regions of the planetesimal disk \citep{malhotra1995, gomes1997}. A way out to avoid this
eccentricity excitation in a resonance sweeping scenario is to consider
corotation resonances instead of libration resonances \citep{ward-canup2006}. In order
to accomplish that one must consider an eccentric perturbing body, in our case
Neptune. Thus I consider a scenario based on the classical Nice model \citep{tsiganis-a-2005, gomes-a-2005, gomesetal2018} in
which the planets, including Neptune, acquire fairly excited orbits during
their close encounter phase. In this scenario, at some point in the planetary
orbital evolutions, the close encounters phase comes to an end and the planets
experience a residual migration with a circularization of their orbits. In the
simulations here presented I assume that the planets have just ended their mutual
close encounters phase (or are about to end) and start their final evolution to their current orbits due to the remaining perturbing planetesimals. In Section 6 I present some results of numerical integrations of planets and massive planetesimals that may give some indication of what kind of disk may be consistent with the assumptions here made for the final phase of planetary migration represented by the synthetic model. 
In this manner I consider the semimajor axes for the giant planets as $a_N=25$ au, $a_U=17$ au, $a_S=9.35$ au and $a_J=5.2$ au.
Jupiter is considered in its
present orbits since it migrates much less than the other giant planets and its small
migration will have roughly no influence in the mechanisms presented in this
work. Uranus and Saturn initial  semimajor axes are based on mean ratios of
their semimajor axes variations to Neptune's
obtained in numerical integrations with a massive planetesimal
disk \citep{gomesetal2018}.  The initial eccentricities of Neptune and Uranus are respectively $0.3$ and $0$, chosen so as to
make them near but not in a a close encounter regime. Other orbital elements were
chosen as the current ones, except for Neptune's mean longitude which was chosen
randomly from $0^{\circ}$ to $360^{\circ}$. Neptune's  mean longitude is the only
orbital element that will be different for each integration and its random choice
will be sufficient to yield quite different evolutions for the integrations. 
The null eccentricity for Uranus is just
a trick to start integrations in order to yield more effective results, in order to avoid deep close encounters between the ice planets from the beginning which may cause the ejection of one of the ice planets. Although artificial, it is not an unrealistic supposition since  in planetary instability migration models Uranus and Neptune experience expressive oscillations of their eccentricities and either of them may be temporarily quite low, what the synthetic model in fact reproduces since Uranus eccentricity is immediately raised by Neptune's high eccentricity.
From those initial
conditions I start numerical integrations of the four giant planets including a
synthetic force on Neptune and Uranus to account for the residual migration and
circularization of their orbits. This force is constructed by applying the
following accelerations on the planets. 

\begin{equation}
A =  K\;\exp{-t/\tau}\;(1+C\,\cos{\lambda})
\end{equation}

\noindent where $A$ is the absolute value of an acceleration applied  in the direction of
the planet's velocity, $K$, $\tau$ and $C$ are constants and $\lambda$ is the planet's mean
longitude. The constant $K$ is defined by $\tau$ and the initial ($a_i$) and final ($a_f$) semimajor axis
of the planet through:

\begin{equation}
K = - {G M \over \tau} ({1 \over a_f} - {1 \over a_i})
\end{equation}

\noindent where $G$ is the gravitational constant and $M$ is the Sun's mass. $\tau$ is
the timescale of the exponential semimajor axis evolution and the constant $C$ controls the variation
of the planet's eccentricity. The values of $\tau$ and $C$ were chosen to mimic
the evolution of the planets' semimajor axes and eccentricities just after the
close encounter phase, as observed in some numerical integration done for previous works \citep{gomesetal2018}.
They are $\tau = 1.37 My$ and and $C=-15$ for most of the planetary evolution. In the beginning (up to $1.4$ My) I use a smaller (in absolute value)
$C = -2$ to account for a time when Neptune still keeps a moderately high eccentricity and guides
the resonance sweeping. When Neptune's eccentricity is lower than $0.1$ I also use a damping variation of $C$
to account for a smooth circularization of the planet's orbit. This is given by:

\begin{equation}
C = C_0 \, (e_P/0.1)^2
\end{equation}

\noindent where $e_P$ is the planet's eccentricity and $C_0$ is the old value of $C = -15$.

The planetesimals are considered massless in the outer border of a planetesimal
disk. I consider this outer part of the disk from $a=39.5$ au
to $42.5$ au and refer to it
hereafter as the Outer Border Disk (OBD). Semimajor axes for the planetesimals are
chosen uniformly on that range. The
eccentricities are chosen randomly in the range 0 to 0.05 and the inclinations
also randomly from $0^{\circ}$ to $1^{\circ}$ with respect to the invariant plane. The eccentricities and inclinations are chosen in a
somewhat ad-hoc manner but aiming at giving a small excitation due to an expected moderate
past perturbation by the excited planets during their close encounters phase.
It is important to note that in many cases of numerical integrations with planets
and a massive planetesimal disk the position from where Neptune starts its final free-of-encounters
migration is attained quite abruptly from a smaller semimajor axis that would only slightly perturb the OBD.
Other orbital elements were chosen randomly from $0^{\circ}$ to $360^{\circ}$.
The OBD is supposed to be the outermost part of the planetesimal disk. The OBD's total mass 
can be approximated to zero in the simulations since it is supposed to perturb
 the planets in a negligible manner.
We also assume that whatever the planetary dynamics
during the close encounter phase was, it did not greatly disturb the OBD.
The inner portion of the disk (below $39.5$ au) is not here considered and is supposed
to have influenced the dynamics of the planets during the close encounters phase
and must be mostly responsible to form the hot population although some of the outermost part of 
this inner portion may also partially contribute to the ECP. The
inner border of the OBD at $39.5$ au was chosen in order to
associate it to Neptune's initial semimajor axis at $25$ au which yields the 2:1 MMR
with that planet at $\sim 39.7$ au. 

\begin{figure}
\centering
\includegraphics[scale=.57]{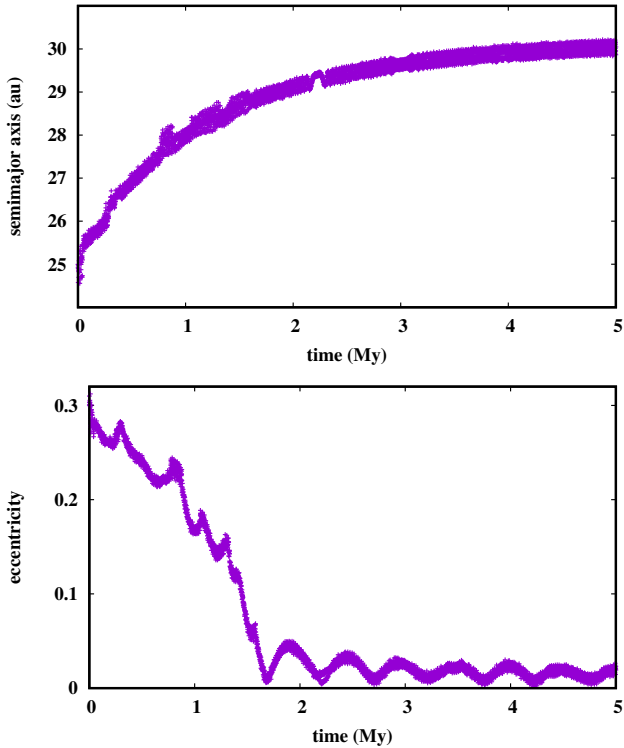}
\caption{Evolution of Neptune's semimajor axis and eccentricity for one of the simulations based on the synthetic model}
\label{fig-tae}
\end{figure}

The choice of random orbital longitudes for Neptune's initial orbit yields
sufficient randomness to the evolution of the pair of ice planets. So although
Neptune and Uranus are initially in non-crossing orbits, in many cases they
attain close encounter orbits in the beginning of their evolution. Some of
these evolutions end with one of the ice planets being ejected from the solar
system. Moreover the final semimajor axes of Uranus and Neptune can differ
from their real actual ones from up to 2 au. This is expected since the synthetic
accelerations were constructed based on a non perturbed orbit. Since I am interested in the final
orbital distribution of the planetesimals started in the OBD when the
planets have their current semimajor axes,
either (1), in the case Neptune stops before 30.1 au, I stretch the
integration for another $1$ My applying synthetic accelerations on Uranus and Neptune so that they
migrate linearly (here with no influence in the eccentricities) and  stop at their current semimajor axes, or
(2), in the case Neptune stops past $30.1$ au, I consider the orbital distribution
of planetesimals at the time when Neptune's mean semimajor axis was at $30.1$ au. Figure \ref{fig-tae}
shows an example of the evolution of Neptune's semimajor axis and eccentricity for one of the integrations.

\begin{figure}
\centering
\includegraphics[scale=.57]{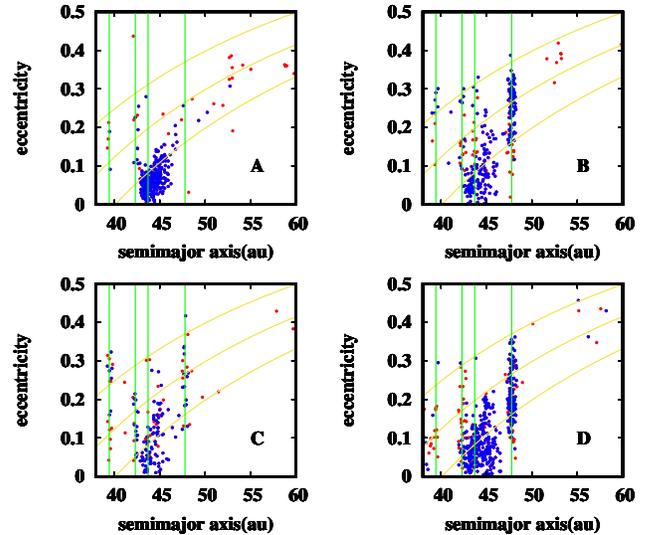}
\caption{Distribution of semimajor axes and eccentricities of planetesimals transported
to the CCKB from four different simulations. Inclinations lower than $4^{\circ}$ are
represented by blue dots, otherwise red dots. The vertical lines stand for the MMRs with Neptune, 
respectively, from left to right, the 3:2, 5:3, 7:4 and 2:1. The curved lines define constant perihelion
distances $q=30$ au, $q=35$ au and $q=40$ au.}
\label{fig-ae}
\end{figure}

\section{Main Results}

The final distribution of the planetesimals initially at the OBD
is finely dependent on the particular evolution of the ice planets, particularly on
Neptune, whereas the planets evolution is finely dependent on the initial
mean longitude of Neptune. In general the planetesimals park at three main
regions: (1) the CCKB, (2) the DCP and (3) in MMRs with Neptune, mostly the 2:1,
3:2, 5:3 and 7:4. It can be quite hard to select the best orbital distribution of the ECP
obtained by the simulations by comparing it
to the real objects' orbital distribution belonging to the ECP as defined by their colors and shown in Fig. \ref{fig-br}. One would have to
compare distributions of semimajor axes, eccentricity and inclination in the
CCKB, ratio of number of objects in the resonances and the DCP to those in the CCKB
, etc. I chose to only compare (1) the semimajor axis distribution in the
CCKB, the ratio of the number of objects in the DCP to the number of objects in
the CCKB and the ratio of the number of objects in the 2:1 MMR with Neptune to
those on the CCKB. More detailed comparisons would complicate too much a problem
that is particularly influenced by observational bias. It must be noted that the 
number of reddish objects in each of the considered subpopulations is influenced
by the desire of the observer to obtain the photometry of that particular object
and this is an important source of bias.
After submitting the final results to the above mild constraints I also submitted 
them to visual inspections and the
best of them were chosen to have their evolution continued to $4.5$ Gy. Four of
these case are shown in Fig \ref{fig-ae}.

Although none of the four cases shows a perfect comparison with the real data
all of them show objects in all subpopulations (CCKB, DCP and MMRs with Neptune).
The very specific orbital distribution of the ECP is finely dependent
on the orbital evolutions of the giant planets and more specifically on Neptune.

The number of planetesimals implanted in the CCKB is just a fraction of those initially
in the OBD. The cases depicted in Fig. \ref{fig-ae} are associated to a fraction 
of the OBD deposited in the CCKB ranging from $1$ to $5$\%. 
From a total mass in the CCKB estimated at $3\times10^{-4} M_{\oplus}$ \citep{fraser-a-2014} to $10$ times as that
\citep{nesvornyetal2020} we can reckon a mass in the
OBD ranging from $0.006 M_{\oplus}$ to $0.3 M_{\oplus}$.
This larger mass may be more compatible 
for the formation of the CCKB objects from streaming instability scenarios that form a 
relatively larger total mass in CCKB objects. On the other hand, the largest mass estimated for the OBD ($0.3 M_{\oplus}$) if extended to $30$ au with the same density as for the OBD would entail a total mass beyond $30$ au of around $1 M_{\oplus}$ which would not be high enough to bring Neptune to the edge of the planetesimal disk \citep{gomesetal2004}. Even for a density of the disk beyond $30$ au to the inner edge of the OBD around five times as large as the density at the OBD, which would imply a total of $5 M_{\oplus}$ beyond $30$ au, Neptune would hardly migrate farther than $30$ au. In Section 6, I present results from numerical integrations of the primordial planets and a disk of massive planetesimals where this question is farther discussed.

\begin{figure}
\centering
\includegraphics[scale=.55, angle=-90]{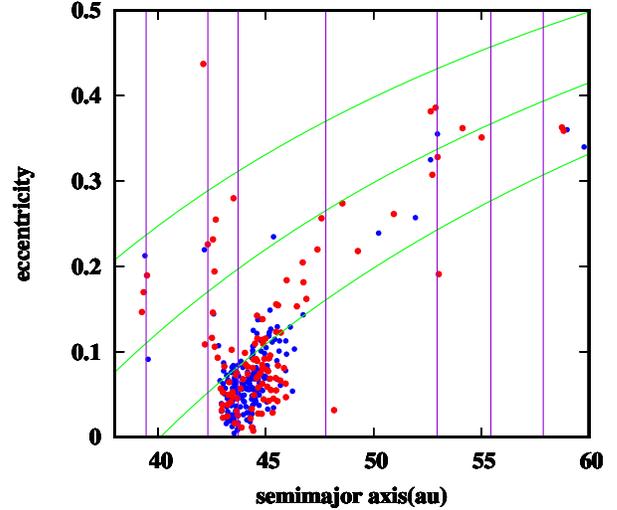}
\caption{Distribution of semimajor axes and eccentricities for the  case in panel A of Fig. \ref{fig-ae}
when, like Fig. \ref{fig-ae}, the outer border of the OBD is at $42.5$ au (blue circles) and for
the outer border at $41.5$ au (red circles) Vertical lines stand for the nominal position of 
the main MMRs with Neptune, from left to right the 3:2, 5:3, 7:4, 2:1, 5:2, 8:3 and 3:1. The curved lines define constant perihelion
distances $q=30$ au, $q=35$ au and $q=40$ au.}
\label{fig-ob}
\end{figure}

The outer border of the OBD was chosen in a somewhat ad-hoc manner, 
only making sure that all particles in the OBD would be transported. But it may not be necessary
that the disk border is exactly at 42.5 au. With the results here obtained we can check for other possible
outer borders for the OBD, just excluding the planetesimals that had their semimajor axes
 above a given value from the final results. In the examples considered here there is not 
much difference in the final distribution of planetesimals in the ECP when we consider a lower 
border for the OBD. A typical example can be noticed in Fig. \ref{fig-ob} where I plot again the
distribution of planetesimals in the ECP for the case of panel A of Fig. \ref{fig-ae}
in the case with the disk border at $42.5$ au (blue)
and at $41.5$ au (red). In this figure, the number of red dots is $136$ and the total number of particles is $328$, 
which correspond to fractions of implanted planetesimals from the shorter and longer OBD 
(with respectively $3333$ and $5000$ planetesimals) respectively
0.041 and 0.066. The fraction of planetesimals implanted in the ECB from
the OBD is usually smaller for an OBD outer border at $41.5$ than at $42.5$.

\begin{figure}
\centering
\includegraphics[scale=.5]{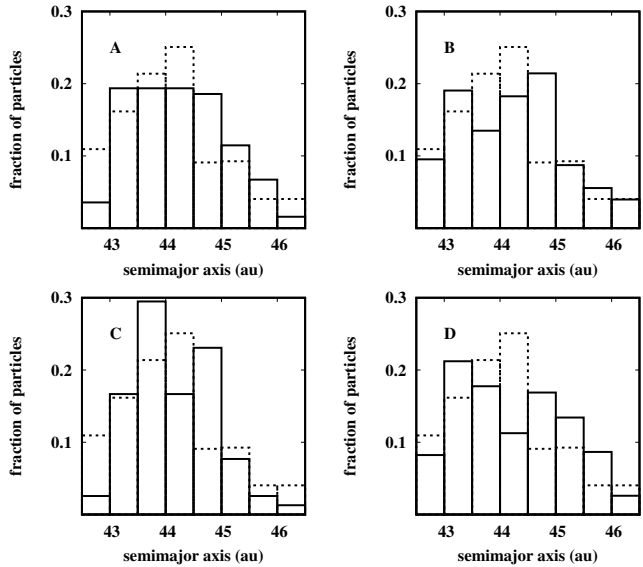}
\caption{Histograms of the distribution of semimajor axes for the same simulations as 
those depicted in Fig. \ref{fig-ae}. The dashed lines stand for the real case}
\label{hist-a}
\end{figure}

Figure \ref{hist-a} shows the histograms of semimajor axes of all planetesimals implanted
in the CCKB from the results depicted in Fig. \ref{fig-ae}, in which the labels A-D correspond 
to the same case as in Fig. \ref{fig-ae}. In dashed line, the histogram for the real CCKB is presented for 
comparison. Again, although in no case there is a perfect agreement, we can conjecture that the real distribution of CCKB 
objects are due to a very specific Neptune's evolution. On the other hand, 
we can see that there is always a bin of semimajor axes for any case (except maybe for case A) in which there is a 
maximum in the number of planetesimals.

\section{Transporting Mechanisms to the ECP}

I consider separately in the following subsections the transporting mechanisms for 
three distinct subpopulations of the ECP, the CCKB, the DCP and the resonant population.

\subsection{The CCKB}

\begin{figure}
\centering
\includegraphics[scale=.45]{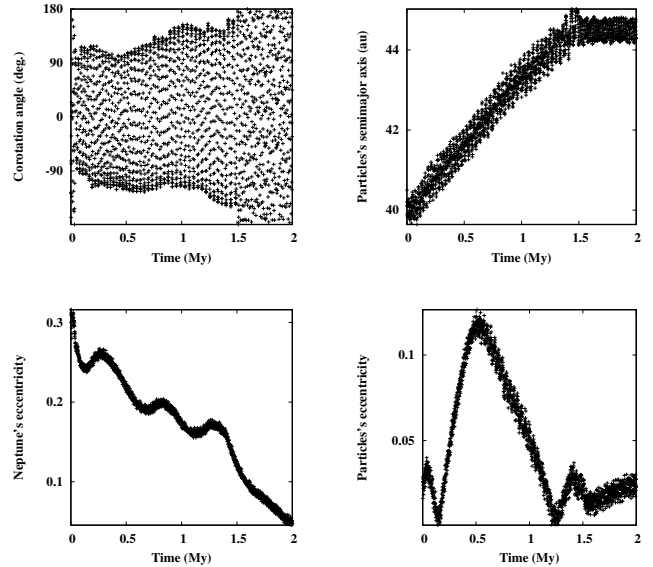}
\caption{Example of the evolution of orbital elements of a planetesimal initially
trapped in the 2:1 corotation MMR with Neptune and released from the resonance at the
CCKB} 
\label{fig-cckb-corr}      
\end{figure}

This is the most important part of the ECP. The main mechanism responsible for
the formation of the CCKB is commented in Section 3 as the motivating reason
for this work. An eccentric Neptune
captures planetesimals into its 2:1 corotation resonance. When this planet
migrates it conveys planetesimals outwards with it and when Neptune's orbit
is around $28$ au and fairly circular several planetesimals are released from the resonance and
implanted in the CCKB region around $44$ au. Of course, the final distribution of
orbits in the CCKB is quite sensitive to the very peculiar evolution of
Neptune's orbit, as depicted in Fig. \ref{fig-ae}. A typical evolution of a planetesimal
that experiences the main mechanism that conveys planetesimals to the CCKB is
shown is Fig. \ref{fig-cckb-corr}. The planetesimal is captured into the 2:1 corotation resonance
with Neptune at the beginning of its evolution, as indicated by the evolution
of the corotation angle ($2 \lambda_P-\lambda_N-\varpi_N$). At some point, when
Neptune's eccentricity is around $0.12$ the resonance is broken and the particle
is parked with $a \sim 44.2$ au and $e \sim 0.03$. It is noteworthy that, contrary to the
libration resonance evolution, in which the planetesimal's  eccentricity
experience a monotonic increase, in the corotation case, there may be
eccentricity increases due to Neptune's own high eccentricity but it is not
monotonic. It shows an oscillatory variation and at the point of resonance break the
eccentricity can be low enough as expected for a member of the CCKB.
Inclination evolutions are not shown. They are kept always low as expected. They mostly
reflect their initial values.
During this kind of planetesimal evolution there is never a close encounter
between it and Neptune, as expected, preserving in this manner the binary
feature of the CCKB objects. 

\begin{figure}
\centering
\includegraphics[scale=.43]{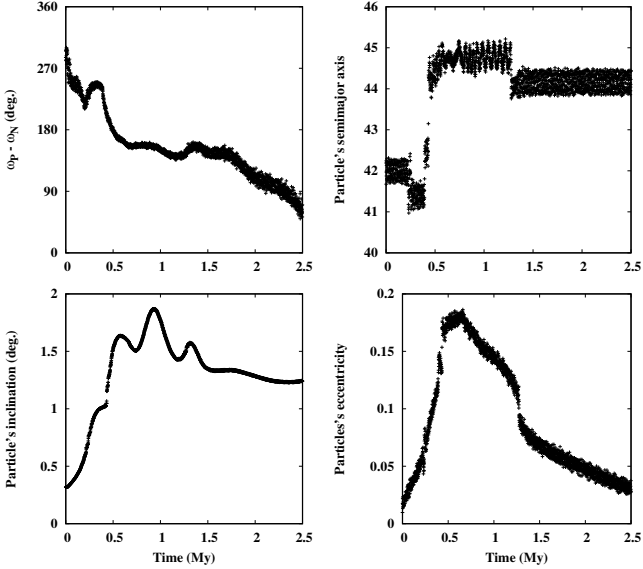}
\caption{Example of the evolution of orbital elements of a planetesimal that was
scattered by Neptune and eventually deposited in the CCKB} 
\label{fig-cckb-clap}      
\end{figure}

\begin{figure}
\centering
\includegraphics[scale=.5]{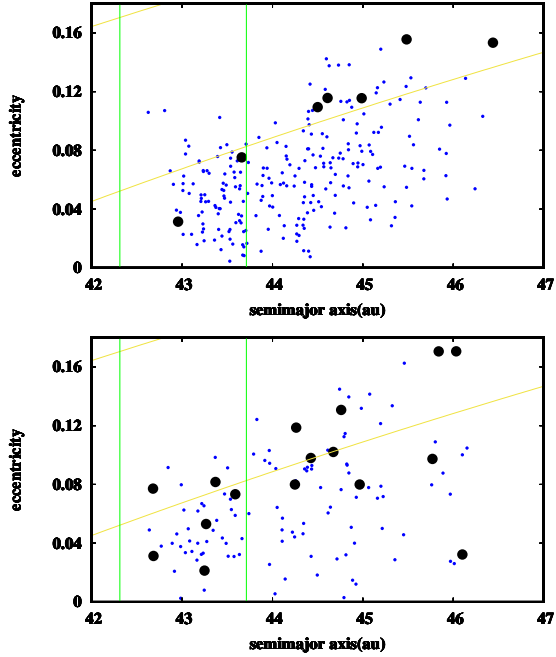}
\caption{Two examples of the distribution of semimajor axes and eccentricities coming from
panel A of Fig. \ref{fig-ae} (upper panel) and panel B of Fig. \ref{fig-ae} (lower panel)
where planetesimals that were at some time closer than $2$ au from Neptune are represented by
large black circles}
\label{cl}
\end{figure}

Yet, there is another mechanism that was found to
occur in several cases of planetesimals deposited in the CCKB. Fig. \ref{fig-cckb-clap}
depicts a
typical such case. It is like the typical case of planetesimals transported to
the DCP and more details of this mechanism will be explained in Section 5.2. But
the complete mechanism includes a fast semimajor axis variation due to close
encounter perturbations with Neptune, an escape from this regime as Neptune's
eccentricity decreases followed by a decrease of the particles's eccentricity
which will be explained with more detail in the following subsection. In most
integrations that yielded a nice CCKB distribution at the end of the migration phase, the prevailing
mechanism is the former one explained above. But the latter one also appears
with non negligible frequency and in some integrations it can reach about 30\%
of the cases. The problem with this mechanism is that it is expected to destroy the
binary feature of the CCKB objects by coming too close to Neptune \citep{parker-kavelaars2010}. But
contrary to one's expectation this is not quite true. For the cases depicted in Fig. \ref{fig-ae}
I redid the integrations computing at every integration step the distance of each planetesimal to Neptune
and I output this distance and the time whenever a planetesimal was less than $2$ au far from Neptune.
In the example shown in Fig. \ref{fig-cckb-clap}, I found that the distance of the planetesimal to 
Neptune was never smaller than $2$ au.
Fig. \ref{cl} shows the distribution of semimajor axes and
eccentricities for the cases in panel A and B of Fig. \ref{fig-ae}, showing only the CCKB and
highlighting the planetesimals that got closer than $2$ au from Neptune by a
larger black circle.  
Case B (lower panel) is the one with more instances of close encounters.
For this case, there are $19$ planetesimals in $126$ (thus $\sim 15$\%) that got closer
than $2$ au from Neptune during their evolutions, all of them
for less than $2$ Ky. For the same case, $10$ planetesimals got closer than $1$ au, all of them
 for less than $210$ years
during their evolutions. Although a detailed
analysis of the evolution of CCKB-like  binary systems perturbed by Neptune during
the evolutions here described is out of my scope, the distances of the planetesimals to Neptune above calculated
suggest with good confidence that most objects transported to the CCKB will
preserve their binary feature.

\subsection{The Distant Cold Population}

\begin{figure}
\centering
\includegraphics[scale=.4]{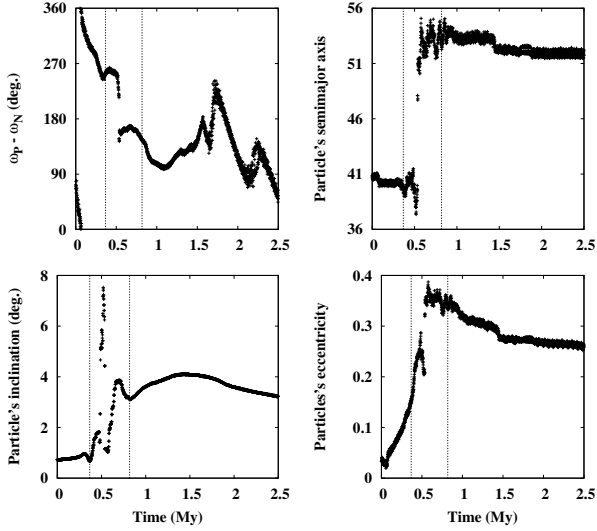}
\caption{Evolution of orbital elements of a planetesimal that was implanted
in the DCP}
\label{fig-ecp-exemplo}
\end{figure}

The second mechanism for transporting planetesimals to the CCKB is the one
responsible to fill the DCP region. Figure \ref{fig-ecp-exemplo} shows the evolution of orbital
elements of a typical case. The typical behavior includes three main phases:
(1) by the effect of an eccentric Neptune, the planetesimal's eccentricity
increases until (2) a phase of close encounters with Neptune starts when the
planetesimal's semimajor axis increases until (3) by Neptune's eccentricity
decrease the close encounter phase ceases and the planetesimal's eccentricity
decreases to a safe value that corresponds to a stable orbit for the planetesimal. The
phase of potential close encounters is delimited by two vertical lines, where
$d<0$.  What deserves an explanation is the planetesimal's eccentricity decrease
after the close encounter phase. This is due to a still strong secular
influence of a still nearby Neptune on the planetesimal and the fact that
$\varpi_P-\varpi_N$ is in the right phase to yield the planetesimal's eccentricity
decrease, which can be seen in the  left-upper panel of Fig. \ref{fig-ecp-exemplo}. 
In fact, to its lowest order, the secular variation of the planetesimal's eccentricity is given by:

\begin{equation}
\dot e = -{1 \over n a^2 e}\, {d R \over d \varpi} \\
\end{equation}

\begin{equation}
R = - {1 \over 4}\, G\, m\, L\, e\, e_N\, \cos{(\varpi-\varpi_N)}
\end{equation}

\noindent where $n$, $a$, $e$ and $\varpi$ are respectively the mean motion, semimajor axis,
eccentricity and longitude of the perihelion. The index $N$ stands for Neptune whereas no index
stands for the planetesimal. $R$ is the disturbing function, $G$ is the gravitational constant, $m$
is Neptune' mass and $L$ a Laplace coefficient
depending on the semimajor axes of the planetesimal and Neptune. Thus we can write:

\begin{equation}
\dot e = - C_+ \sin{(\varpi-\varpi_N)}
\end{equation}

\noindent where $C_+ = (G\, m\, L \, e_N)/(4\, n\, a^2)$ is a positive constant (depending on $e_N$). Thus when $\varpi-\varpi_N$ is between
$0^{\circ}$ and $180^{\circ}$, $\dot e < 0$. Moreover just after the close encounter
phase when $e_N$ is still large, the effect on $\dot e$ is likewise large.

Figure \ref{fig-ecp}  shows the
variation of $\varpi_P-\varpi_N$, where the index $P$ stands for a planetesimal,
$q_P$ and $q_P-Q_N$, where $q_P$ is the perihelion of the planetesimal
and $Q_N$ is Neptune's aphelion. They
refer to 
planetesimals that parked in the DCP at $10$ My from the integration
depicted in panel A  of Fig. \ref{fig-ae}. It can be noticed that the perihelia
of these planetesimals experience an increase during the time $\varpi_P-\varpi_N$ is
in the right range. Of course after a planetesimal is free from close
encounters with Neptune, the behavior of its eccentricity will depend on the
specific phase of $\varpi_P-\varpi_N$ and those that are in the right phase
to have their eccentricity decreased are just a fraction of all that were
scattered by Neptune. Fig. \ref{fig-ecp-exemplo-cont} shows the continuation of the evolution of
$\varpi_P-\varpi_N$ and the eccentricity of the same planetesimal as that of Fig. \ref{fig-ecp-exemplo}.
When the planetesimal and Neptune are far enough $\varpi_P-\varpi_N$ circulates faster
and the eccentricity starts a periodic variation around its relatively low
eccentricity. 

\begin{figure}
\centering
\includegraphics[scale=.5]{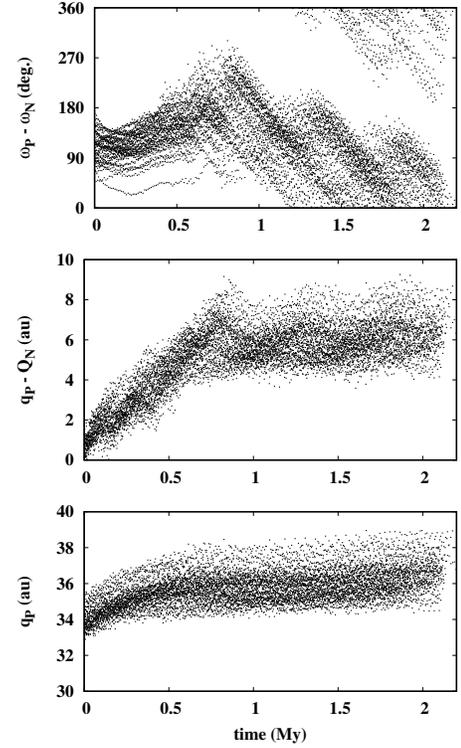}
\caption{Evolution of the orbital elements of several planetesimals that were
implanted in the DCP: planetesimals' perihelia, the difference of planetesimals'
perihelia to Neptune's aphelion and the difference of planetesimals' longitudes of the perihelion
 to Neptune's 
longitude of the perihelion.}
\label{fig-ecp}
\end{figure}

Now some words are in order about the close encounter history of the DCP planetesimals as
formed by the above mechanism. Figure \ref{hist-dclap} shows a histogram of smallest close
encounter distances from Neptune of the planetesimals that became a member of the
DCP at $3$ Gy, from the run depicted in panel A of Fig \ref{fig-ae}. This can be
compared with Fig. 2 of \cite{parker-kavelaars2010}. Studying
the final effect on putative binaries of the encounters that
preceded the implantation of these objects into the DCP is out of the scope of this work.
On the other hand, the inventory of binary TNOs 
with semimajor axis in the range $50$ au to $60$ au and 
that have their B-R magnitude measured as greater than $1.58$ is still small for an accurate statistics.
Moreover whatever the mechanisms that were responsible for
the implantation of red objects into the scattered / detached population, it is expected
that these mechanisms included episodes of close encounters with Neptune, thus 
one should not expect the preservation of binary systems as a rule.

\begin{figure} 
\centering
\includegraphics[scale=.4]{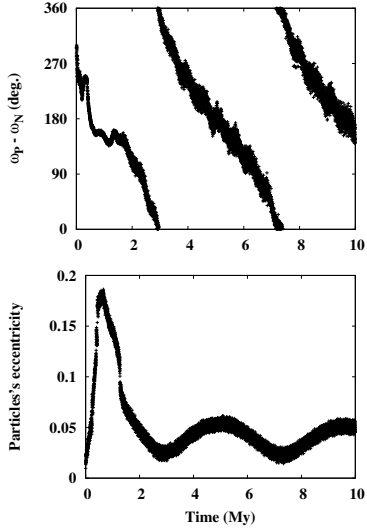}
\caption{Continuation to $10$ My of the evolution of the eccentricity
and the difference of the planetesimal's longitude of the perihelion to
Neptune's, for the same case as in Fig. \ref{fig-ecp-exemplo}}
\label{fig-ecp-exemplo-cont}
\end{figure}

\begin{figure}
\centering
\includegraphics[scale=.4]{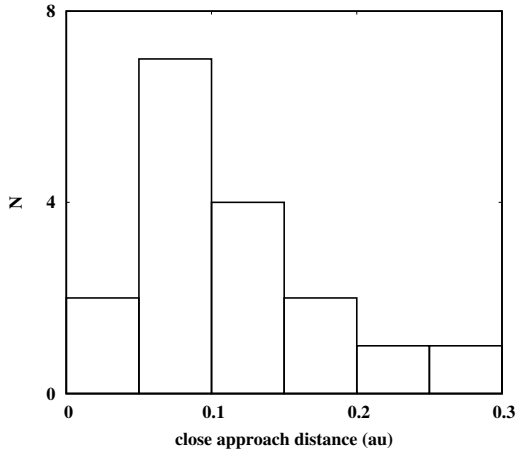}
\caption{histogram of smallest close encounter distances from Neptune of the planetesimals that became
 a member of the DCP at $10$ My, from the run depicted in panel A of Fig \ref{fig-ae}}
\label{hist-dclap}
\end{figure}

\subsection{The Resonant Population}

\begin{figure}
\centering
\includegraphics[scale=.45]{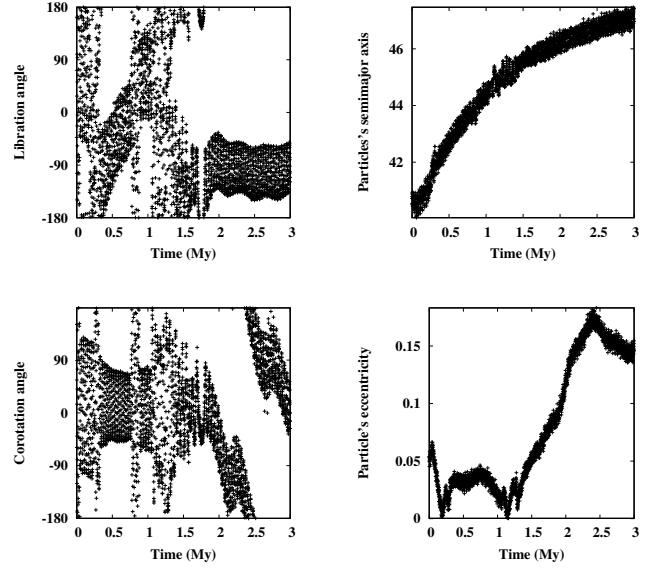}
\caption{Evolution of orbital elements of a planetesimal that was trapped in the 2:1 MMR with Neptune}
\label{fig-r12}
\end{figure}

The mechanism which is the easiest to explain the filling of a resonant region with ECP objects
is that associated with the 2:1 resonance. Figure \ref{fig-r12} shows a typical case.
The peculiar behavior in this case is depicted by the evolution of the 2:1
resonant angle, in which the active resonant angle alternates from a
corotation to a libration and this change occurs at about $1.7$ My. This example
is didactic since it shows that the corotation and libration resonances  are
not well separated. While the corotation resonance is active, the libration
angle shows a kind of libration with a varying libration center, whereas after
$1.7$ My the opposite is true. This is associated to the fact that $\varpi_P-\varpi_N$
has a relatively slow variation and this on the other hand is likely what drives the specific
variation of the particle's eccentricity while in corotation
resonance with Neptune. The total number of planetesimals trapped into the 2:1 MMR
with Neptune as well as the ratio of this number to the number of planetesimals implanted in the CCKB
depends sensitively on the very final evolution of Neptune as suggested
by Fig. \ref{fig-ae}. Thus the evolution depicted in Fig. \ref{fig-r12} suggests that understanding
the process by which planetesimals are either released in the CCKB or switch from corotation
to libration resonance is fundamental to possibly constrain this final evolution of
Neptune so as to produce the actual relative distribution of planetesimals in the CCKB and the 2:1 MMR.
The eccentricities attained by the planetesimals in the 2:1 MMR with Neptune by the proposed mechanism,
which is approximately between $0.15$ and $0.3$, implies an original trapping by Neptune into the libration
resonance when the planetesimals are between $42$ au and $44$ au with eccentricities from $0$ to $0.1$
considering their monotonic variation induced by the exchange of angular momentum and energy with the
migrating Neptune by which the planetesimals are being swept in resonance 
\citep{gomes1997}.

The other important resonances with Neptune are the 3:2, 5:3 and 7:4. As for
these last two, the trapping into these resonances usually occurs as a secondary
process after the planetesimal has been transported some way out by the same
process that creates the CCKB. As for the 3:2 resonance, there is always a
close encounter with Neptune and the planetesimal eventually enters the
3:2 resonance from a scattering orbit by a resonance sticking mechanism. In some cases, there is a residual
migration associated with Neptune's residual migration. The 3:2 resonance is
the less typical according to the scenario proposed in this work. It does not
mean that it will be less populated by ECP particles since I chose $39.5$ au as
the inner border of the OBD in order to explain the formation of the CCKB
and DCP. There must be other reddish particles originally with $a<39.5$ and some of these
must be affected by the 3:2 resonance with a migrating Neptune.
But then this process will depend significantly on the very
specific evolution of Neptune before the assumed beginning used in the
simulations here presented. This scenario that can produce Plutinos would
need another approach which is beyond the scope of this work.

\section{Does the Synthetic Model have a real Counterpart?}

\begin{figure}
\centering
\includegraphics[scale=.55]{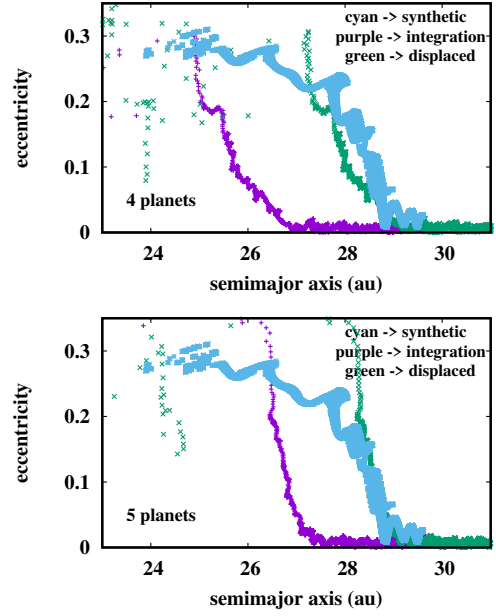}
\caption{Comparison of the evolution of Neptune's semimajor axis and eccentricity
for the synthetic case that generated the example of panel A in Fig. \ref{fig-ae}
with Neptune's orbital elements from two numerical simulations of the planets in a massive
planetesimal disk, one case with four planets and another one with five planets. Neptune's
orbital elements are with respect to the barycenter of the system including the Sun and the other
three planets inside Neptune's orbit.}
\label{fig-compar-4-5planets}
\end{figure}

\begin{table*}
\begin{center}
\caption{Initial orbital elements for the planets}
\label{elspl}
\begin{tabular}{lccc}
\\
\tableline\tableline
 Planet & semi-major axis (au) & eccentricity & inclination (deg.)\\
\tableline
5-Planets Model \\
Jupiter	  &    5.60141  &    0.02357  &    0.02627  \\
Saturn 	  &    7.32016  &    0.08123  &    0.06867  \\
Core1  	  &   10.28600  &    0.11279  &    0.11508  \\
Core2  	  &   12.29610  &    0.06361  &    0.06063  \\
Core3  	  &   16.56300  &    0.02534  &    0.04085  \\
\\
4-Planets Model \\
Jupiter	  &    5.41097  &    0.04089  &    0.05080  \\
Saturn 	  &    8.63125  &    0.01907  &    0.05676  \\
Core1  	  &   11.30040  &    0.03608  &    0.09070  \\
Core2  	  &   14.93020  &    0.00629  &    0.03500  \\
\tableline
\end{tabular}
\end{center}
\end{table*}

The model presented in Section 3 was constructed aiming at a specific result, 
bringing the outermost part of the primordial disk of planetesimals to fill
the present CCKB and the rest of the ECP. A fair question that imposes is:
Can real evolutions of the present giant planets (and possibly other cores that
were ejected from the Solar System) migrating in a planetesimal disk yield the 
conditions for Neptune to experience the specific evolution presented in Section 3
that can spread the original OBD onto the ECB? In order to answer this question I did
several numerical integrations of the giant planets perturbed by a planetesimal disk.
I considered both a four-planets model and a five-planets model \citep{nesv-2011, nesv-morb-2012}. 
Table \ref{elspl} gives the initial
orbital elements of the planets. The planetesimal disk for the 4-planets model 
was considered with a total
mass chosen randomly for each integration 
between $25 M_{\oplus}$ to $40 M_{\oplus}$, an inner border at $13$ au and an outer border at $35$ au,
and for the 5-planets model, the mass was $25 M_{\oplus}$ to $35 M_{\oplus}$, the inner border at $17$ au
and the outer border also at $35$ au.
These conditions were chosen based on previous integrations \citep{gomesetal2018} aiming at 
parking the planets near their present orbits but also aiming at allowing Neptune
to experience a final evolution similar to that presented in Section 3. In this respect 
stretching the disk to $35$ au is important in order to allow Neptune to have its eccentricity
decreased along with a small but non negligible semimajor axis increase. For this to be
accomplished when Neptune is eccentric its aphelion must invade a planetesimal populated region
otherwise the orbital circularization will not concomitantly be followed by a semimajor axis
increase.

\begin{figure}
\centering
\includegraphics[scale=.55]{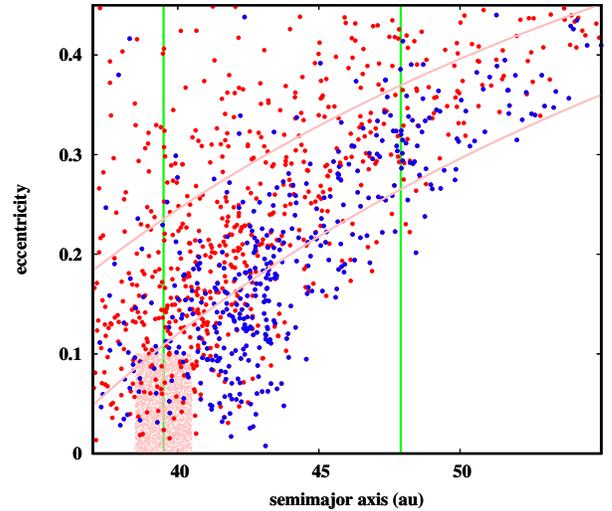}
\caption{Initial (pink dots) and final semimajor axis and eccentricities distribution
of planetesimals in an outer border disk that was perturbed by the planets in the
numerical simulation with five planets as depicted in Fig. \ref{fig-compar-4-5planets}, lower panel.
Blue circles stand for inclinations below $5^{\circ}$ and red circles otherwise. Vertical
lines stand for the position of the 3:2 and 2:1 resonance with Neptune. Other lines
stand for the semimajor axes and eccentricities that yield $q=a_N$ and $q=a_N+5$ au.}
\label{fig-numinteg-5pls-ae}
\end{figure}

Figure \ref{fig-compar-4-5planets} shows the variation of the semimajor axis with the eccentricity of Neptune
for two cases, one from the 4-planets model and the other one for the 5-planets model.
In both cases they are compared to a synthetic evolution on Neptune, the one associated
with panel A of Fig. \ref{fig-ae}. In both cases I also plot the numerical integration case
displaced in semimajor axis so as to superimpose it to the synthetic model. This is just an
artifice to better
show that the variation of semimajor axis and eccentricity imposed by the synthetic
model has a real counterpart except possibly for the position of Neptune when its final
decrease of eccentricity takes place. This  behavior occurs for both models but is more
frequent for the 4-planets one. This is possibly caused by the fact that the initial
semimajor axis for the outermost planet in the 4-planets model is not so far as for
the 5-planets model. This makes us guess whether a 6-planets model might give still better results
in this sense.  This is a way to shift the outermost planet farther outwards and keep the planets
close enough to trigger a planetary instability scenario that can yield an eccentric enough
Neptune which is required for the present model. An investigation of a six-planets model
can be useful to check for that possibility and and this is planned as a future work. 

With the initial orbital elements as shown in Table \ref{elspl} and the disk masses as indicated above, Neptune's average semimajor axis at the end of the integrations at $50$ My was $26.7$ au and $28.6$ au, respectively for the 4-planets and the 5-planets model. Allowing for $1 - 2$ au extra migration for Neptune at $4.5$ Gy (based on some integrations that were extended to $4.5$ Gy) we can estimate that the mass chosen for the planetesimal disk was reasonable so as not to let Neptune migrate too far \citep{gomesetal2004}. The planetesimal disks extension and mass entail a total mass beyond $30 au$ (and to $35$ au) between $6.9 M_{\oplus}$ and $9.6 M_{\oplus}$ which seems consistent with the estimated mass in the OBD of $0.3 M_{\oplus}$ at most, as indicated in Section 4. It must be noted that migration models that yield eccentric planetary orbits produce final semimajor axis for Neptune systematically 
smaller than for the case of models where Neptune sweeps a cold planetesimal disk. 

Figure \ref{fig-numinteg-5pls-ae} shows the distribution of semimajor axes and eccentricities of massless particles
started with the semimajor axis and eccentricities depicted in the same figure with pink dots. These initial
conditions for the planetesimals represent an OBD displaced inward with respect to that considered in the synthetic 
simulations since, as shown in Fig.  \ref{fig-compar-4-5planets} lower panel, Neptune's eccentricity decrease
starts from a lower semimajor axis as compared to the synthetic simulations.
Larger dots stand for the same particles at $50$ My of the numerical integration including the perturbing 
planets that evolved according to the 5-planets numerical simulation
that yielded Neptune's semimajor axis and eccentricity evolutions as shown in the bottom
panel of Fig. \ref{fig-compar-4-5planets}. Blue dots stand for particles with the inclination $I < 5^{\circ}$
and red dots otherwise. Although in several aspects different from the real distribution of 
the ECP, this figure however shows the main characteristics of the formation of the ECP: a CCKB
between the 3:2 and 2:1 MMR with Neptune, the DCP beyond the 2:1 resonance and a few particles
at the 2:1 MMR with Neptune. The main difference is in the location of the CCKB which is shifted towards
the 3:2 MMR with Neptune and consequently farther from the 2:1 MMR. It is also transported
on a smaller range from the initial OBD as compared to the results from the synthetic model. There is also an excess of 
dynamically hot particles, but this can be blamed at the small time span for which the integration
was developed. Most high eccentricity particles must be shifted out of the CCKB after $4.5$ Gy.
 A nice result shown by this figure is the formation of low inclination, 
high perihelion DCP particles, a result in better accordance with the real observed DCP than
the DCP obtained by the synthetic model. The shortcomings described above with respect
to the position of the CCKB comes from the fact that Neptune's decrease in eccentricity 
occurs at lower semimajor axis and there is a longer residual migration of Neptune. 
More simulations can be done to check for the possibility of a better evolution of Neptune,
including a 6-planets model.

\section{Discussion and Conclusions}

Cold Classical Kuiper Belt objects are today consensually considered as having an
in situ formation. It is also consensus that CCKB objects have a specific physical
characteristic evidenced by their reddish colors. This particular feature claims for
the need of a common origin for all reddish TNOs which argues for a mechanism that may
both implant the CCKB in its present location but also other reddish objects (Extended
Cold Population) on their present locations.
This fact motivated the present work in which an alternative scenario
 for the appearance of the Cold Classical Kuiper Belt is presented. 
In this scenario, the objects pertaining to the CCKB were transported a short range from the
outer border of a primordial planetesimal disk. The same mechanism that implanted the CCKB on its
present location would also be responsible for the implantation of objects with similar color
characteristics as scattered / detached objects with moderately low inclinations. The mechanism
is based on a classical Nice model in which Neptune acquires a high eccentricity and
experiences a simultaneous orbital circularization and residual migration that are
responsible to moderately convey the outer planetesimal disk outwards, keeping their
relatively low eccentricities and low inclinations. The planetesimals are transported by a
2:1 corotation resonance sweeping process with Neptune and the planetsimals' implantation is accomplished
by the eventual decrease of Neptune's eccentricity and the planetesimals release from the resonance.

An important byproduct of this mechanism is the formation of a relatively low inclination
population of scattered and detached objects beyond but not far from the 2:1 MMR with Neptune. There
are real objects occupying that region and they are not easily if at all explainable by
a local formation of the CCKB. In fact, the mechanism here presented requires an episode
of an eccentric Neptune as opposed to a scenario where Neptune is never or just slightly eccentric
\citep{nesvorny2015}. A scenario with a local formation of the CCKB and an episode of an eccentric
Neptune was also checked by this author in previous numerical simulations \citep{gomesetal2018}
and no DCP was detected in the data from those simulations. This is expected since the
relative excitation of the CCKB induced by an eccentric Neptune is provoked by pure secular
dynamics with no close encounter episodes that could at least moderately scatter the planetesimals
from the primordial CCKB. Also in a scenario where Neptune always or mostly migrates with low eccentricity
\citep{nesvorny2015} there is always a 'migrating' scattered population with Neptune but no mechanism to
detach them to larger perihelia keeping low inclinations, since there is a void of important MMRs with Neptune just past
the 2:1 MMR with Neptune and classical mechanisms that produce the detached objects require
important MMRs with Neptune and a fairly high planetesimal's  inclination to allow for the coupled
MMR Kozai mechanism that increases the planetesimal's perihelion distance with a simultaneous 
inclination increase.

Even though streaming instability theories have given a large step towards explaining the
formation of large planetesimals from a small amount of mass in pebbles, it is still questionable whether
the right conditions in the primordial solid-gas disk was attained to form the objects in the present
CCKB. The short range transport scenario here presented may eventually show to be compatible to more
realistic conditions for the primordial gas-solids disk, since the objects as large as those found in the CCKB 
primordially in the OBD would sum to a total mass of 20-100 times the present CCKB mass.

Finally, although the scenario here presented dispenses a local formation for the CCKB it is however
not incompatible with an in situ formation of part of the CCKB. In this case a fraction of the CCKB would
have a local formation and another part would have been transported. There might be then some small 
differences in the physical characteristics of those two subpopulations. Possibly the loca1
formed objects would belong to the 
reddest classes. It is however noteworthy that three of the detached low inclination objects above mentioned,
2001 UR163, 1995 TL8 and  2002 GZ31, have their B-R magnitudes in the range $1.75$  to $1.97$, two of them belonging to the
RR class, thus already pushed towards very red objects.

%\bibliographystyle{cas-model2-names}

%\bibliography{cas-refs}

\bibliographystyle{aasjournal}

\bibliography{bib}

\end{document}